Letter to the Editor

# Water vapor toward starless cores: the *Herschel* view[⋆]

P. Caselli[1,2], E. Keto[3], L. Pagani[4], Y. Aikawa[5], U.A. Yıldız[6], F.F.S. van der Tak[7,8], M. Tafalla[9], E.A. Bergin[10],
B. Nisini[11], C. Codella[2], E.F. van Dishoeck[6,12], R. Bachiller[9], A. Baudry[14], M. Benedettini[18], A.O. Benz[13],
P. Bjerkeli[17], G.A. Blake[19], S. Bontemps[14], J. Braine[14], S. Bruderer[13], J. Cernicharo[20], F. Daniel[20], A.M. di Giorgio[18],
C. Dominik[21,22], S.D. Doty[23], P. Encrenaz[4], M. Fich[24], A. Fuente[25], T. Gaier[26], T. Giannini[11], J.R. Goicoechea[20],
Th. de Graauw[7], F. Helmich[7], G.J. Herczeg[12], F. Herpin[14], M.R. Hogerheijde[6], B. Jackson[7], T. Jacq[2], H. Javadi[26],
D. Johnstone[15,16], J.K. Jørgensen[27], D. Kester[7], L.E. Kristensen[6], W. Laauwen[7], B. Larsson[28], D. Lis[29], R. Liseau[17],
W. Luinge[7], M. Marseille[7], C. M$^c$Coey[24,30], A. Megej[46], G. Melnick[3], D. Neufeld[31], M. Olberg[17], B. Parise[32],
J.C. Pearson[26], R. Plume[33], C. Risacher[7], J. Santiago-García[34], P. Saraceno[18], R. Shipman[7], P. Siegel[26],
T.A. van Kempen[3], R. Visser[6], S.F. Wampfler[13], and F. Wyrowski[32]

(Affiliations can be found after the references)

**ABSTRACT**

*Aims.* Previous studies by the satellites SWAS and Odin provided stringent upper limits on the gas phase water abundance of dark clouds ($x$(H$_2$O) <7×10$^{-9}$). We investigate the chemistry of water vapor in starless cores beyond the previous upper limits using the highly improved angular resolution and sensitivity of *Herschel* and measure the abundance of water vapor during evolutionary stages just preceding star formation.
*Methods.* High spectral resolution observations of the fundamental ortho water (o-H$_2$O) transition (557 GHz) were carried out with the Heterodyne Instrument for the Far Infrared onboard *Herschel* toward two starless cores: Barnard 68 (hereafter B68), a Bok globule, and LDN 1544 (L1544), a prestellar core embedded in the Taurus molecular cloud complex. Detailed radiative transfer and chemical codes were used to analyze the data.
*Results.* The rms in the brightness temperature measured for the B68 and L1544 spectra is 2.0 and 2.2 mK, respectively, in a velocity bin of 0.59 km s$^{-1}$. The continuum level is 3.5±0.2 mK in B68 and 11.4±0.4 mK in L1544. No significant feature is detected in B68 and the 3 $\sigma$ upper limit is consistent with a column density of o-H$_2$O $N$(o-H$_2$O) < 2.5×10$^{13}$ cm$^{-2}$, or a fractional abundance $x$(o-H$_2$O) < 1.3×10$^{-9}$, more than an order of magnitude lower than the SWAS upper limit on this source. The L1544 spectrum shows an absorption feature at a 5 $\sigma$ level from which we obtain the first value of the o-H$_2$O column density ever measured in dark clouds: $N$(o-H$_2$O) = (8±4)×10$^{12}$ cm$^{-2}$. The corresponding fractional abundance is $x$(o-H$_2$O) ≃ 5×10$^{-9}$ at radii > 7000 AU and ≃2×10$^{-10}$ toward the center. The radiative transfer analysis shows that this is consistent with a $x$(o-H$_2$O) profile peaking at ≃10$^{-8}$, 0.1 pc away from the core center, where both freeze-out and photodissociation are negligible.
*Conclusions.* *Herschel* has provided the first measurement of water vapor in dark regions. Column densities of o-H$_2$O are low, but prestellar cores such as L1544 (with their high central densities, strong continuum, and large envelopes) appear to be very promising tools to finally shed light on the solid/vapor balance of water in molecular clouds and oxygen chemistry in the earliest stages of star formation.

## 1. Introduction

The abundance of water vapor in dark clouds is a crucial missing quantity in our understanding of astrochemistry and gas-grain interactions. Before the launch of the Submillimeter Wave Astronomy Satellite (SWAS), pseudo-time-dependent gas phase chemical models predicted water fractional abundances ($x$(H$_2$O) ≡ $n$(H$_2$O)/$n$(H$_2$), where $n(i)$ is the number density of species $i$) around 10$^{-7}$ (e.g. Lee et al. 1996). SWAS observations of the o-H$_2$O ($J_{KK'}$:1$_{10}$-1$_{01}$) line toward the low-mass starless cores B68 and Oph D place stringent upper limits on the abundance of water vapor in cold gas of $x$(H$_2$O) < 3×10$^{-8}$ and <8×10$^{-9}$ in B68 and Oph D, respectively (Bergin & Snell 2002). The Odin satellite provided an even lower upper limit toward the protostellar core Cha-MMS1 (<7×10$^{-9}$; Klotz et al. 2008). Water in dark clouds is mostly in the form of ice mantles covering dust grains ($x$(H$_2$O)$_{\rm ice}$ ≃ 10$^{-4}$; e.g., Whittet & Duley 1991).

The gas-phase chemical models needed revision. The inclusion of dust grains in the chemistry and the freeze-out of oxygen onto grain surfaces significantly improved the agreement with observations (e.g. Bergin et al. 2000, Viti et al. 2001, Roberts & Herbst 2002 Melnick & Bergin 2005, Aikawa et al. 2005). Hollenbach et al. (2009) also pointed out the importance of illumination by the external far-ultraviolet (FUV) radiation field. They showed that in clouds of constant density illuminated by the local interstellar field ($G_0$ = 1), the water abundance can reach a peak of about 10$^{-7}$ only in a range of local visual extinctions $A_V$ between 1 and 4 mag. At $A_V$ < 1, H$_2$O molecules are photodissociated, whereas at $A_V$ > 4 mag the water is mostly in solid form. Only in-between these ranges does a combination of photodesorption from grain mantles and gas phase chemistry allow a relatively high water vapor abundance.

In spite of our improved understanding of the oxygen chemistry, the water problem is not yet settled. There are two main

---

[⋆] *Herschel* is an ESA space observatory with science instruments provided by European-led Principal Investigator consortia and with important participation from NASA.

reasons for this. First, detailed radiative transfer codes have shown that line trapping and absorption of the dust continuum challenge the determination of the amount of water vapor in the dense clouds (Poelman et al. 2007). Second, oxygen in dense gas is poorly constrained because non-thermal desorption processes can release adsorbed O atoms from grain surfaces that would otherwise be transformed into water ice (Caselli et al. 2002a). There is indeed evidence of abundant atomic oxygen toward the molecular cloud L1689N (Caux et al. 1999) and W49N (Vastel et al. 2000). Dick et al. (2010) inferred extremely low collisional cross-sections between $H_2O$ and $H_2$ at temperatures below 40 K, so that in cold clouds the ortho and para ground-state lines of water may only be observed in absorption, even in rich water vapor environments. So, are starless cores dry? The question remains open, hindering a detailed understanding of the oxygen chemistry and budget in dense regions. The higher angular resolution and much higher sensitivity of *Herschel* (Pilbratt et al. 2010), compared to Odin and SWAS (20 to 40 better, respectively, for the 557 GHz line), give us a unique opportunity to further investigate the water vapor content.

In this Letter, we present the first observations in the direction of two well-studied starless cores, the Bok globule B68 and preliminary results about the prestellar core L1544, embedded in the Taurus molecular cloud complex. These two objects have been extensively studied; their density, temperature, and velocity profiles have been measured and successfully modeled (e.g. Tafalla et al. 1998, Alves et al. 2001, Evans et al. 2001, Caselli et al. 2002b, Hotzel et al. 2002, Bergin et al. 2006, Broderick et al. 2007, Crapsi et al. 2007, Kirk et al. 2007, Keto & Caselli 2010). They are representative of the two different classes of starless cores: thermally subcritical (B68) and supercritical (L1544; Keto & Caselli 2008) and they are both far from active sites of star formation. Therefore, they are the ideal sites for studying the water content along the lines-of-sight with dense molecular material not yet affected by protostellar feedback. Observations of the centrally concentrated and contracting L1544 provide insight into the water content just before a protostar is born, thus helping to set the initial chemical conditions in the process of star formation.

## 2. Observations

As part of the WISH (Water In Star-forming regions with *Herschel*) key project (van Dishoeck et al., in prep.), we observed the o-$H_2O$ ($1_{10}$-$1_{01}$) line at 556936.0020 MHz toward B68 (RA(J2000) = $17^h22^m38^s.20$, Dec(J2000) = $-23°49'54''.0$, $V_{LSR}$ = 3.4 km s$^{-1}$) and L1544 (RA(J2000) = $05^h04^m17^s.21$, Dec(J2000) = $25°10'42''.8$, $V_{LSR}$ = 7.2 km s$^{-1}$) with *Herschel* HIFI (de Graauw et al. 2010), on 2010 March 20. The dual beam switch (dbs) mode, with the continuum optimization and fast chop options was used. The wide-band spectrometer (wbs) and the high resolution spectrometer (hrs) were used simultaneously in both H and V polarizations. A beam efficiency of 0.75 was used to convert the antenna temperature scale into main beam brightness temperature ($T_{mb}$).

The individual spectra were reduced using the *Herschel* interactive processing environment (hipe, version 3.0.1; Ott 2010). A final analysis was performed in CLASS[1]. The H and V spectra were analyzed individually to check for any discrepancy in the calibration. No significant difference was noticed, so the two spectra were averaged. The wbs spectra were smoothed at the nominal 1.1 MHz resolution (0.59 km s$^{-1}$) and the rms noise level was 2.0 mK for B68 and 2.2 mK for L1544 (1.4 times lower than the rms reached in the hrs spectra after Gaussian smoothing to the same wbs resolution, the expected difference between the two backends). Only wbs spectra will be considered here.

[1] CLASS is the continuum and line-analysis single-dish software within the GILDAS package available at http://www.iram.fr/IRAMFR/GILDAS

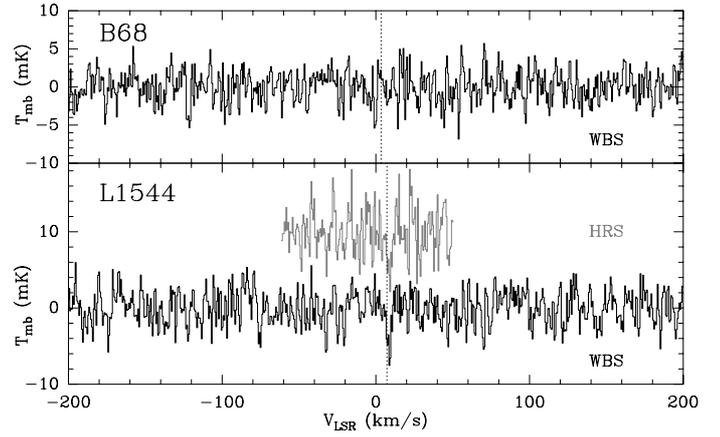

**Fig. 1.** Wbs spectra (black) of B68 and L1544, and hrs spectrum (grey) of L1544, centered on the o-$H_2O$($1_{10}$-$1_{01}$) frequency. Baselines have been subtracted. The hrs spectrum has been Gaussian smoothed at the wbs resolution and shifted by +10 mK. The dotted vertical lines are the LSR velocities measured from high density tracers. Note the absorption feature in the wbs and hrs spectra of L1544.

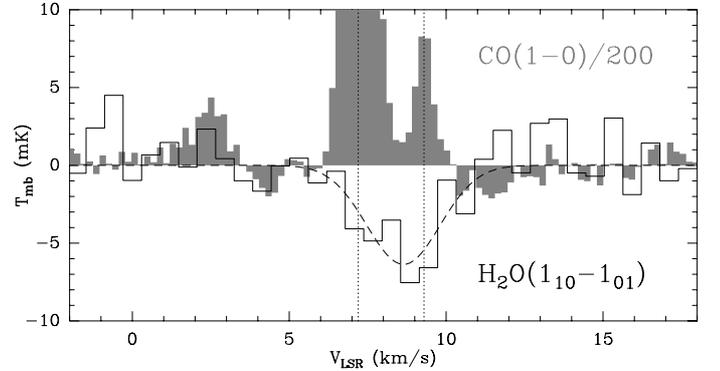

**Fig. 2.** Zoom-in of Fig. 1 for L1544. The empty histogram is the $H_2O$($1_{10}$-$1_{01}$) wbs spectrum and the shaded histogram is the CO(1-0) spectrum taken with FCRAO (Tafalla et al., in prep.). The CO line shows two velocity components, one centered on the core velocity, 7.2 km s$^{-1}$ ($T_{mb}$ = 12 K) and the other on 9.3 km s$^{-1}$ (dotted lines). Note the good match. The dashed curve is the Gaussian fit to the absorption feature.

## 3. Results

Dust continuum emission in the B68 and L1544 spectra is observed at a level of 3.5±0.2 and 11.4±0.4 mK (1.6±0.1 and 5.1±0.2 Jy), respectively. Figure 1 shows the two spectra after baseline subtraction. No significant feature is seen in the direction of B68, whereas a faint absorption line is present towards L1544, also visible in the hrs spectrum (grey histogram). Figure 2 is a zoom-in of the wbs L1544 spectrum in Fig. 1 (empty histogram), plotted with the CO(J:1-0) line observed with the Five College Radio Astronomy Observatory (FCRAO) (shaded histogram; Tafalla et al., in prep.). The o-$H_2O$ feature is broad and covers the two velocity components of CO(1-0): the first is centered on the dense core velocity (7.2 km s$^{-1}$;

Caselli et al. 2002a), whereas the other (also observed in CO(2-1), but not in high density tracers) peaks at $9.29\pm0.03$ km s$^{-1}$. *This is the first evidence of water vapor in the direction of a cold core.* Interestingly, the main absorption feature is coincident with the weaker component of the CO (1-0) spectrum, suggesting that most of the absorption is not associated with the dense core material. In the following, we first derive the water vapor abundance in the two starless cores assuming homogeneous conditions. We then present our detailed radiative transfer analysis.

## 3.1. Basic estimates

### 3.1.1. B68

An upper limit to the fractional abundance of o-H$_2$O can be obtained from the B68 spectrum (Fig. 1). We used the RADEX code (van der Tak et al. 2007), with input parameters appropriate for B68 of kinetic temperature $T_k$ = 10 K, line width $\Delta v$ = 0.5 km s$^{-1}$ (based on CO observations and accounting for the different molecular masses), and H$_2$ volume densities $n$(H$_2$) = $10^4$ cm$^{-3}$. To reproduce the observed 3 $\sigma$ upper limits of 6 mK, with the wbs resolution, the required column density of o-H$_2$O is $N$(o-H$_2$O) = $4.2\times10^{14}$ cm$^{-2}$. Within the *Herschel* beam (FWHM = 39″), $N$(H$_2$) $\simeq 2\times10^{22}$ cm$^{-2}$ (e.g. Alves et al. 2001), thus, for B68, $x$(o-H$_2$O) < $2.1\times10^{-8}$, consistent with the SWAS upper limit (Bergin & Snell 2002). However, this analysis implies that: (i) the dust continuum emission at 557 GHz (539 $\mu$m) is negligible, and (ii) there is a uniform physical (excitation) structure. This is not correct (see Sect. 3.2). If the line is in absorption, it cannot be stronger than 1.3 $\sigma$ deep because the continuum is 1.3 $\sigma$ above the 2.73 K background, and the line is therefore undetectable. Thus, no upper limit can be given.

### 3.1.2. L1544

The absorption feature toward L1544 allows us to measure, for the first time, the column of absorbing water in front of the far-infrared (FIR)-emitting core material. A Gaussian fit to the feature (dashed curve in Fig. 2) gives an integrated intensity $I$ = $-18.5\pm4.0$ mK km s$^{-1}$, line width $\Delta v$ = $2.7\pm0.6$ km s$^{-1}$, $T_{mb}$ = $-6\pm2$ mK, and $V_{LSR}$ = $8.6\pm0.3$ km s$^{-1}$ (identical, within the errors, to the hrs spectrum fit results). From the continuum level and $T_{mb}$, we obtain an optical depth $\tau$ = $0.6\pm0.4$. With $\tau$ and $\Delta v$, it is straightforward to obtain $N$(o-H$_2$O) = $(8\pm6)\times10^{12}$ cm$^{-2}$. To arrive at $x$(o-H$_2$O), $N$(H$_2$) is needed. Assuming that CO (1-0) and H$_2$O($1_{10}$-$1_{01}$) trace the same material (as suggested by the similar velocities), we estimate $N$(H$_2$) from the CO (1-0) line integrated intensity adopting an X-factor of $2\times10^{20}$ cm$^{-2}$ K$^{-1}$ km$^{-1}$ (e.g. Pineda et al. 2008), which gives $N$(H$_2$) = $3\times10^{21}$ cm$^{-2}$. Thus, $x$(o-H$_2$O) = $(5\pm4)\times10^{-9}$. We note that our $N$(H$_2$) estimate is about 20 times lower than the value found using dust continuum emission at 1.2 mm within the *Herschel* beam (Ward-Thompson et al. 1999), and the 539 $\mu$m continuum observed by *Herschel*, assuming a dust opacity at 230 GHz $\kappa_{230}$ = 0.009 cm$^2$ g$^{-1}$, a dust emissivity spectral index $\beta$ = 2 (e.g. Schnee et al. 2010), and a dust temperature of 8 K within the *Herschel* beam (Crapsi et al. 2007): $N$(H$_2$) $\simeq$ $7\times10^{22}$ cm$^{-2}$. This discrepancy is expected given that CO is mostly frozen-out within the central $\simeq$7000 AU (e.g. Caselli et al. 1999), where the bulk of the mm and FIR continuum is emitted. Using the $N$(H$_2$) value derived from the continuum, $x$(o-H$_2$O) $\simeq 2\times10^{-10}$, suggesting that the water vapor abundance is very low toward the core center, like CO. The $x$(o-H$_2$O) values derived here are averages along the line of sight. Detailed radiative transfer modeling and more sensitive observations are needed to place constraints on the abundance profile (see Sect. 3.2).

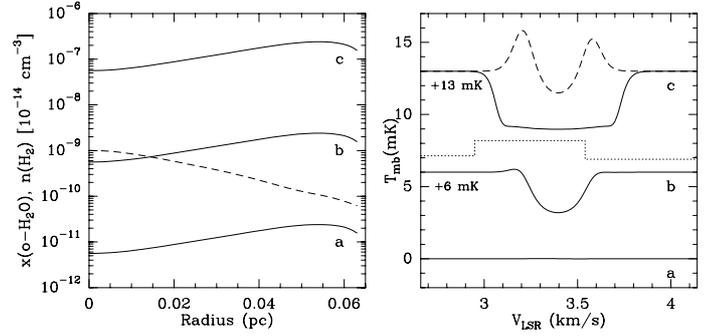

**Fig. 3.** *Left panel*. Abundance profiles of o-H$_2$O adopted in the radiative transfer MOLLIE for B68. The dashed curve is the H$_2$ volume density profile of B68 (see KC08). *Right panel*. MOLLIE results for the three different $x$(o-H$_2$O) profiles in the left panel. For displaying purposes, the continuum level has been subtracted. Simulated **b** and **c** spectra have been shifted by 6 and 13 mK, respectively. DDP collisional coefficients have been used. For comparison, the dashed curve shows the result of adopting PMG rates. The dotted histogram is the wbs spectrum from Fig. 1 after adding +6 mK to compare with the simulated spectrum **b**.

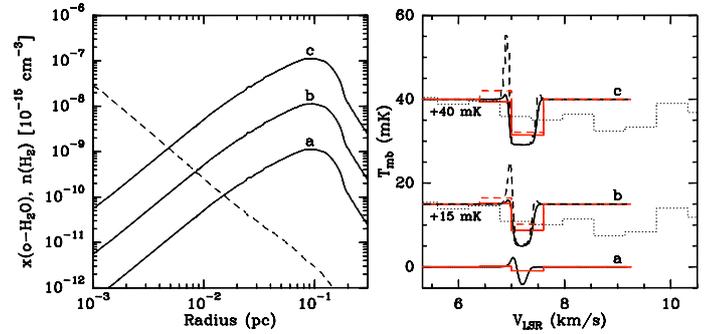

**Fig. 4.** Same as Fig. 3 but for L1544. In the right panel, the dotted histogram is the wbs spectrum from Fig. 1 after adding +15 and +40 mK to compare with the simulated spectra **b** and **c**. The red solid histogram is the black curve (profile obtained using DDP rates) smoothed to 0.6 km s$^{-1}$. The red dashed histogram is the dashed black curve (profile obtained using PMG rates) smoothed to 0.6 km s$^{-1}$. Note that the core velocity component can be reproduced by models **b** and **c**.

## 3.2. Radiative transfer including dust emission

Keto & Caselli (2008, 2010, hereafter KC08 and KC10) used the MOLLIE radiative transfer code and successfully reproduced observed line intensities and profiles in B68 and L1544. From the comparison between observed and simulated spectra, KC08 derived temperature and density profiles of the two starless cores, as well as their velocity structure. The analysis of the present *Herschel* HIFI data is based on the KC10 models, which were modified to include water vapor in the code. Two different sets of collisional coefficients are used in the modeling: those from Dick et al. (2010, hereafter DDP) and those from Phillips et al. (1996, hereafter PMG). At temperatures below about 40 K, DDP provide the lowest rates, whereas PMG give the highest rates available, so together they represent the two extreme values present in the literature. The rates of Dubernet et al. (2009) will be considered in an upcoming paper.

We considered abundance models by Aikawa et al. (2005, hereafter AHRC), Hollenbach et al. (2009), and KC10 (freeze-out, a generalized desorption from Roberts et al. 2007, and photodissociation, but no detailed gas phase and surface chemistry) and we plot those of KC10. The more sophisticated models of AHRC and Hollenbach et al. (2009) predict different abundances that also differ from those of KC10. We will investigate these differences when more sensitive data will be available. Example $H_2O$ abundance profiles for the two sources are shown in Figs. 3 and 4, left panels, where the density profiles can also be found. Curves labeled **a**, **b**, and **c** refer to different peak abundances which are compared with the present data.

The dust continuum emission is taken into account by MOLLIE and the observed fluxes at 539 $\mu$m are reproduced with the dust properties found by KC08 and KC10. MOLLIE results for the three water abundance profiles are shown in Figs. 3 and 4, right panels. For B68, the upper limit is consistent with all three curves, the only exception being curve **c**, *if DDP rates are correct*. The o-$H_2O$ column density found by integrating curve **b** along the line of sight, and convolving with the *Herschel* beam, is $N(o-H_2O) = 2.5 \times 10^{13}$ cm$^{-2}$, about 20 times lower than the upper limit found in Sect. 3.1.1. For L1544, curves **b** and **c** in Fig. 4 can explain the absorption features at the dense core velocity (the higher velocity component will be studied with the new *Herschel* data for this source). With the current sensitivity and spectral resolution, it is impossible to distinguish between the two sets of curves and collisional coefficients. For model **c**, the corresponding o-$H_2O$ column density within the *Herschel* beam is $N(o-H_2O) = 3.3 \times 10^{13}$ cm$^{-2}$, about 4 times higher than that measured from the absorption feature (Sect. 3.1.2), so it can probably be ruled out. Using model **b**, we find that $N(o-H_2O) = 3.3 \times 10^{12}$ cm$^{-2}$, which is consistent with the measured value.

## 4. Conclusions

We have presented *Herschel* HIFI observations of o-$H_2O$ ($1_{10}$-$1_{01}$) toward two starless cores embedded in quiescent environments (B68 and L1544). We have interpreted and discussed these data with the help of detailed radiative transfer modeling. In B68, the 3 $\sigma$ upper limit is consistent with a column density of ortho water $N(o-H_2O) < 2.5 \times 10^{13}$ cm$^{-2}$, or a fractional abundance $x(o-H_2O) < 1.3 \times 10^{-9}$, about twenty times lower than the previous upper limit found with SWAS. The L1544 spectrum contains an absorption feature that closely matches the velocity range spanned by the CO (1-0) line. The feature appears to be double peaked (as does the CO line). From the absorption, we have derived the first value of the o-$H_2O$ column density ever measured in dark and cold clouds, $N(o-H_2O) = (8\pm4) \times 10^{12}$ cm$^{-2}$, or a fractional abundance of $x(o-H_2O) \simeq 5 \times 10^{-9}$ in the outer regions traced by CO (radii > 7000 AU), and $\simeq 2 \times 10^{-10}$ toward the core center. From our radiative transfer analysis, we have found that the velocity component associated with the dense core is consistent with an abundance profile of o-$H_2O$ peaking at a value of $\simeq 10^{-8}$, 0.1 pc away from the core center. At this radius, the volume density ($n(H_2) = 5 \times 10^3$ cm$^{-3}$) is not large enough for significant freeze-out to occur and photodissociation has dropped to negligible rates (visual extinction $\leq 2$ mag).

The prestellar core L1544 is more centrally concentrated than B68 and is undergoing gravitational collapse (e.g. Caselli et al. 2002b, van der Tak et al. 2005). The brighter FIR-continuum emission and the larger envelope (L1544 is embedded in the Taurus molecular cloud complex, whereas B68 is an isolated globule) most likely explains the observed absorption. The centrally concentrated and thermally supercritical starless cores, such as L1544 appear to be extremely good candidates for observations of water vapor in dark regions. Thanks to the high sensitivity and angular resolution of *Herschel*, these studies will provide a unique opportunity to finally shed light on the solid/vapor balance of water in molecular clouds and oxygen chemistry in the earliest stages of star formation.

*Acknowledgements.* HIFI has been designed and built by a consortium of institutes and university departments from across Europe, Canada and the United States under the leadership of SRON Netherlands Institute for Space Research, Groningen, The Netherlands and with major contributions from Germany, France and the US. Consortium members are: Canada: CSA, U.Waterloo; France: CESR, LAB, LERMA, IRAM; Germany: KOSMA, MPIfR, MPS; Ireland, NUI Maynooth; Italy: ASI, IFSI-INAF, Osservatorio Astrofisico di Arcetri-INAF; Netherlands: SRON, TUD; Poland: CAMK, CBK; Spain: Observatorio Astronómico Nacional (IGN), Centro de Astrobiologa (CSIC-INTA). Sweden: Chalmers University of Technology - MC2, RSS & GARD; Onsala Space Observatory; Swedish National Space Board, Stockholm University - Stockholm Observatory; Switzerland: ETH Zurich, FHNW; USA: Caltech, JPL, NHSC. HCSS / HSpot / hipe are joint developments by the *Herschel* Science Ground Segment Consortium, consisting of ESA, the NASA *Herschel* Science Center, and the HIFI, PACS and SPIRE consortia.

1. School of Physics and Astronomy, University of Leeds, Leeds LS2 9JT, UK
2. INAF - Osservatorio Astrofisico di Arcetri, Largo E. Fermi 5, 50125 Firenze, Italy
3. Harvard-Smithsonian Center for Astrophysics, 60 Garden Street, MS 42, Cambridge, MA 02138, USA
4. LERMA and UMR 8112 du CNRS, Observatoire de Paris, 61 Av. de l'Observatoire, 75014 Paris, France
5. Department of Earth and Planetary Sciences, Kobe University, Nada, Kobe 657-8501, Japan
6. Leiden Observatory, Leiden University, PO Box 9513, 2300 RA Leiden, The Netherlands
7. SRON Netherlands Institute for Space Research, PO Box 800, 9700 AV, Groningen, The Netherlands
8. Kapteyn Astronomical Institute, University of Groningen, PO Box 800, 9700 AV, Groningen, The Netherlands
9. Observatorio Astronómico Nacional (IGN), Calle Alfonso XII, 3, 28014, Madrid, Spain
10. Department of Astronomy, The University of Michigan, 500 Church Street, Ann Arbor, MI 48109-1042, USA
11. INAF - Osservatorio Astronomico di Roma, 00040 Monte Porzio catone, Italy
12. Max Planck Institut für Extraterrestrische Physik, Giessenbachstrasse 1, 85748 Garching, Germany
13. Institute of Astronomy, ETH Zurich, 8093 Zurich, Switzerland
14. Université de Bordeaux, Laboratoire d'Astrophysique de Bordeaux, France; CNRS/INSU, UMR 5804, B.P. 89, 33271 Floirac cedex, France
15. National Research Council Canada, Herzberg Institute of Astrophysics, 5071 West Saanich Road, Victoria, BC V9E 2E7, Canada
16. Department of Physics and Astronomy, University of Victoria, Victoria, BC V8P 1A1, Canada
17. Department of Radio and Space Science, Chalmers University of Technology, Onsala Space Observatory, 439 92 Onsala, Sweden
18. INAF - Istituto di Fisica dello Spazio Interplanetario, Area di Ricerca di Tor Vergata, via Fosso del Cavaliere 100, 00133 Roma, Italy
19. California Institute of Technology, Division of Geological and Planetary Sciences, MS 150-21, Pasadena, CA 91125, USA
20. Centro de Astrobiología. Departamento de Astrofísica. CSIC-INTA. Carretera de Ajalvir, Km 4, Torrejón de Ardoz. 28850, Madrid, Spain.
21. Astronomical Institute Anton Pannekoek, University of Amsterdam, Kruislaan 403, 1098 SJ Amsterdam, The Netherlands
22. Department of Astrophysics/IMAPP, Radboud University Nijmegen, P.O. Box 9010, 6500 GL Nijmegen, The Netherlands
23. Department of Physics and Astronomy, Denison University, Granville, OH, 43023, USA
24. University of Waterloo, Department of Physics and Astronomy, Waterloo, Ontario, Canada
25. Observatorio Astronómico Nacional, Apartado 112, 28803 Alcalá de Henares, Spain
26. Jet Propulsion Laboratory, California Institute of Technology, Pasadena, CA 91109, USA
27. Centre for Star and Planet Formation, Natural History Museum of Denmark, University of Copenhagen, Øster Voldgade 5-7, DK-1350 Copenhagen K., Denmark
28. Department of Astronomy, Stockholm University, AlbaNova, 106 91 Stockholm, Sweden
29. California Institute of Technology, Cahill Center for Astronomy and Astrophysics, MS 301-17, Pasadena, CA 91125, USA
30. the University of Western Ontario, Department of Physics and Astronomy, London, Ontario, N6A 3K7, Canada
31. Department of Physics and Astronomy, Johns Hopkins University, 3400 North Charles Street, Baltimore, MD 21218, USA
32. Max-Planck-Institut für Radioastronomie, Auf dem Hügel 69, 53121 Bonn, Germany
33. Department of Physics and Astronomy, University of Calgary, Calgary, T2N 1N4, AB, Canada
34. Instituto de Radioastronomía Milimétrica (IRAM), Avenida Divina Pastora 7, Núcleo Central, E-18012 Granada, Spain
35. Université Pierre et Marie Curie, LPMAA UMR CNRS 7092, Case 76, 4 place Jussieu, 75252 Paris Cedex 05, France
36. Observatoire de Paris-Meudon, LUTH UMR CNRS 8102, 5 place Jules Janssen, 92195 Meudon Cedex, France
37. Department of Physics and Astronomy, San Jose State University, One Washington Square, San Jose, CA 95192, USA
38. Laboratoire d'Astrophysique de Grenoble, CNRS/Université Joseph Fourier (UMR5571) BP 53, F-38041 Grenoble cedex 9, France
39. European Southern Observatory, Karl-Schwarzschild-Str. 2, 85748 Garching, Germany
40. Department of Physics, The University of Tokyo, Hongo, Bunkyo-ku, Tokyo 113-0033, Japan
41. Department of Physics and Astronomy, University College London, Gower Street, London WC1E6BT
42. Department of Physics, The University of Tokyo, Hongo, Bunkyo-ku, Tokyo 113-0033, Japan
43. KOSMA, I. Physik. Institut, Universität zu Köln, Zülpicher Str. 77, D 50937 Köln, Germany
44. California Institute of Technology, 1200 E. California Bl., MC 100-22, Pasadena, CA. 91125 USA
45. Experimental Physics Dept., National University of Ireland Maynooth, Co. Kildare. Ireland
46. Microwave Laboratory, ETH Zurich, 8092 Zurich, Switzerland